# Refined-Segmentation R-CNN: A Two-stage Convolutional Neural Network for Punctate White Matter Lesion Segmentation in Preterm Infants

Yalong Liu[1], Jie Li[1], Ying Wang[1], Miaomiao Wang[2],
Xianjun Li[2], Zhicheng Jiao[3], Jian Yang[2], and Xingbo Gao[1]

[1] Lab of Video and Image Processing Systems, School of Electronic Engineering,
Xidian University, Xi'an 710071, China
[2] Department of Radiology, The First Affiliated Hospital of Xi'an Jiaotong University,
Xi'an 710061, China
[3] University of North Carolina at Chapel Hill, Chapel Hill, NC 27599, USA
xbgao@ieee.org

**Abstract.** Accurate segmentation of punctate white matter lesion (PWML) in infantile brains by an automatic algorithm can reduce the potential risk of post-natal development. How to segment PWML effectively has become one of the active topics in medical image segmentation in recent years. In this paper, we construct an efficient two-stage PWML semantic segmentation network based on the characteristics of the lesion, called refined segmentation R-CNN (RS R-CNN). We propose a heuristic RPN (H-RPN) which can utilize surrounding information around the PWMLs for heuristic segmentation. Also, we design a lightweight segmentation network to segment the lesion in a fast way. Densely connected conditional random field (DCRF) is used to optimize the segmentation results. We only use T1w MRIs to segment PWMLs. The result shows that our model can well segment the lesion of ordinary size or even pixel size. The Dice similarity coefficient reaches 0.6616, the sensitivity is 0.7069, the specificity is 0.9997, and the Hausdorff distance is 52.9130. The proposed method outperforms the state-of-the-art algorithm. (The code of this paper is available on https://github.com/YalongLiu/Refined-Segmentation-R-CNN)

## 1    Introduction

Punctate White Matter Lesion (PWML) is a common disease in preterm infants. About 20% of newborns whose gestational age less than 37 weeks suffer from PWML [1]. PWML may result in cerebral palsy and many other neurological disorders [2] if the newborns are not given the artificial intervention in the first time.

At present, we generally believe that the algorithms of feature extraction and segmentation for adult brain tumors are well developed. However, there are many difficulties in the segmentation of PWML compared with that of adult brain tumors. (1)

The immature myelination results in an inversion of MRI contrast compared with adult brain scans [3], so we should perform algorithms on T2w MRIs rather than T1w MRIs. However, the PWML is mainly detected in T1w MRIs. (2) The lesion area is extremely tiny relative to the whole brain area, so the imbalance between positive and negative samples makes it a big challenge for deep learning models. (3) Some of the lesion areas are so small that the final segmentation result may ultimately fail even if there is a little segmentation deviation.

Many attempts have been made to automatically segment PWML. First, Cheng et al. [4] proposed a method based on the threshold. This method can well segment PWMLs in high contrast MRIs. After that, Cheng et al. [5] used a random process to model the neonatal brain MRI, which avoided the assumption that the features of the lesion obeyed the Gaussian distribution. Recently, the method proposed by Mukherjee et al. [6] considered the correlation of pixels in 2D and 3D space and finally improved the performance of PWML segmentation. However, there are much fewer studies on how to keep the robustness of the model on a larger data set.

As we all know, deep learning has been widely used in brain tissue segmentation [7], and brain tumor segmentation [8] in recent years. In this paper, we propose a deep learning method called Refined-Segmentation R-CNN (RS R-CNN) to segment the PWML efficiently. It uses a two-stage method like Mask R-CNN [9] and Faster R-CNN [10] which extract candidate regions first and then process candidate regions. Our model has many advantages compared with other methods. First, the current PWML segmentation algorithms need to remove the skull in MRIs [4-6] while our method does not need to perform any complicated pre-processing. Some of the current algorithms need multi-modality data [4] while our method can efficiently identify the lesion region by only using the T1w MRI. Besides, we use heuristic methods to extract regions of interests (ROIs), so that the surrounding information of ROIs can guide the segmentation progress and lead to more accurate results. Finally, we design a segmentation network module called Lightweight Refined Segmentation Network (LRS-Net), which has the advantages of low resource consumption and high segmentation accuracy. Densely connected conditional random fields (DCRF) is also included in the LRS-Net to optimize its output probability map. Experimental results show that each module of our segmentation algorithm contributes to the performance of the whole model. Also, our algorithm is based on 2D slices rather than 3D voxels for fewer parameters and faster speed. When compared with other existing methods, our model achieves better results in the segmentation of PWMLs.

## 2    Methods

The structure of our proposed method is shown in Fig. 1, which mainly includes the Heuristic Region Proposal Network (H-RPN), the Lightweight Refined Segmentation Network (LRS-Net) and the network head. H-RPN is used to extract feature maps of different layers and generate the ROIs with its surrounding information from the input images. The network head utilizes ROI's surrounding information provided by H-RPN to refine the class and bounding box of the ROI. Then the modified ROIs will be sent to the LRS-Net for lesion segmentation.

Also, we use two independent RoIAligns [9] (RoIAlign1&RoIAlign2) in our model. They can rescale the input feature maps into a new shape by bilinear interpolation. RoIAlign1 rescales the ROIs extracted by H-RPN to $7 \times 7 \times 256$. Then the results of refined class and bounding box can be achieved by the network head according to the rescaled ROIs. The RoIAlign2 will adjust and rescale the ROIs to the size of the input of LRS-Net ($N \times N \times 256$) by the refined results. Then the ROIs will be sent to LRS-Net to get segmentation probability map. The semantic segmentation results of all lesion regions in one slice will be restored to the mask of the original image size according to the location information of the corresponding ROIs. Finally, the semantic segmentation results of the whole image can be obtained.

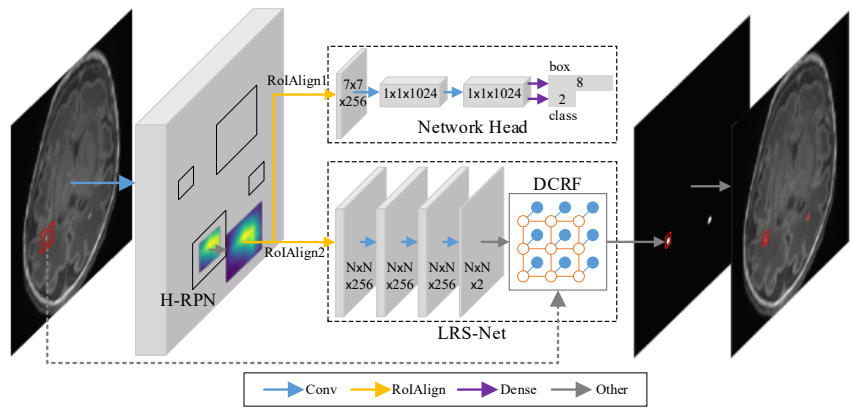

**Fig. 1.** The architecture of the proposed method. It mainly includes the H-RPN, the LRS-Net, and the network head. Blue, yellow and purple arrows respectively indicate convolution, RoIAlign, and Dense connection while the grey arrow represents other processing steps.

## 2.1 Heuristic Region Proposal Network(H-RPN)

The Heuristic Region Proposal Network (H-RPN) consists of a backbone based on ResNet-101 [11] and a Feature Pyramid Network (FPN) [12] as a common RPN. It is used to generate ROIs from the input image. However, we enhance its ability to generate ROIs more efficiently.

Firstly, there are some images with tiny target areas in the PWML data set. The common RPN will get a large number of ROIs which contain few true positive (TP) targets but a large number of false positive (FP) targets by processing this kind of image. Therefore, we enhance the H-RPN by using a rectified linear unit (ReLU) as the activation function of the classifier in H-RPN. The ReLU improves the non-linear ability of the classifier so that the H-RPN can extract more TP regions. Then we reduce the upper limit of candidate region number to 50 in H-RPN, which makes the ratio of positive and negative samples extracted by H-RPN is no longer imbalance.

Secondly, there will be some inevitable deviation when H-RPN intercepts the corresponding ROIs from the last convolutional layer in the backbone. A small deviation of the bounding box is acceptable when the target area is large. However, some lesions are so small that some of the target areas may be outside the bounding box even if there is a small deviation. In this way, it is impossible to segment the entire lesion from such an ROI accurately. In order to alleviate this problem, we proposed a heuristic method. We expand the size of the bounding box to include the surrounding areas of the corresponding ROI in H-RPN. So the second-stage network can obtain more accurate results by the guiding of the surrounding information.

An expansion factor $k\,(k \geq 1)$ is used to expand the bounding box to a new size. For example, when the original bounding box of an ROI is $(y_1, x_1, y_2, x_2)$, then its weight $(w) = x_2 - x_1$ and height $(h) = y_2 - y_1$. So the new bounding box $(y_1^{'}, x_1^{'}, y_2^{'}, x_2^{'})$ can be obtained by Eq. (1).

$$
\begin{aligned}
y_1^{'} &= R\big(y_1 - k \times (h/2)\big), \quad y_2^{'} = R\big(y_2 + k \times (h/2)\big), \\
x_1^{'} &= R\big(x_1 - k \times (w/2)\big), \quad x_2^{'} = R\big(x_2 + k \times (w/2)\big).
\end{aligned}
\tag{1}
$$

Where $R(x)$ represents a round function. In this way, the strong coupling connection between first-stage and second-stage in the deep learning framework will be changed into a weak coupling connection. In other words, we reduce the interference of the prediction error of the bounding box in H-RPN to the second-stage network.

The expansion factor $k$ also affects RoIAlign2. If the height and width of the expanded ROIs exceed that of the original output size of RoIAlign2 ($14 \times 14$), the feature map will be rescaled into a smaller size after the RoIAlign. Some details of the ROIs will be lost in the process which is harmful to the segmentation of the network. So we should adjust the output size of RoIAlign2 to $R(14k) \times R(14) \times 256$.

## 2.2 Lightweight Refined Segmentation Network (LRS-Net)

In the segmentation of PWML, we designed a simple and effective semantic segmentation network called Lightweight Refined Segmentation Network (LRS-Net) which mainly includes the convolution layers and the densely connected conditional random fields (DCRF) [13]. The convolution layers of our segmentation network reduce the number of parameters from 2.6M to 1.2M and get better performance compared with that in Mask R-CNN. As is shown in Fig. 1, the convolution layer of LRS-Net has a simple layered structure. The first layer consists of a $3 \times 3$ convolution with 256 output channels and a ReLU activation. It is used to process the ROIs outputted by RoIAlign2. The input size of the LRS-Net ($N \times N \times 256$) is equal to the output size of RoIAlign2, so $N = R(14k)$. The second layer has the same structure as the first layer. The third layer is the output layer with a $1 \times 1$ convolution and a sigmoid activation.

DCRF is used by several methods as a post-processing method to modify the output of the segmented network and get good results [14, 15]. Considering the sparseness of PWML, we introduce DCRF directly into the interior of the model to optimize the output probability map of the convolution layers. Finally, the segmentation results of all lesions in one slice will be restored to their original positions by the corresponding bounding box information in the network head and H-RPN.

### 2.3 Implementation Details

We implement the proposed method in Keras using the TensorFlow backend. All the models are trained and tested on an NVIDIA Titan X GPU. We use the pre-trained weights of Mask R-CNN on MS-COCO data set and fine-tune our models on the PWML data set. The optimizer is stochastic gradient descent (SGD) with a momentum of 0.9 and a weight decay of 0.0001. The learning rate is 0.001 and will be divided by 2 when it cannot reach a lower validation loss in 10 consecutive epochs. Training epoch is 100. The other loss functions are the same as that in Mask R-CNN. For faster processing, we use Bayesian optimizer [16] which is based on Gaussian process other than random search to adjust the hyper-parameters in DCRF.

## 3 Experiments and Results

The data set enrolled in this study contains 70 subjects in which 704 slices have PWMLs. It is provided by a cooperative hospital with a 3T MRI scanner and confirmed by an experienced neonatal neurologist according to conventional MRI sequences (T1w and T2w). The original image size of one slice is 512 × 512.

We adjust the width of the image by a random deformation coefficient $d(0.85 < d < 1.15)$ and lateral flip the image according to the probability of 0.5 to enhance the data set. We can significantly increase the diversity of the data by doing these measures. Then the images will be cut into the original size (512 × 512), and the intensity distribution will be normalized into zero mean and unit variance.

The train/val/test ratio is 0.7/0.15/0.15. First, we test the impact of different modules on the performance of the baseline model. Then we use *dice similarity coefficient* (DSC), *sensitivity*, *specificity* and *Hausdorff distance* (HD) to compare the performance of our final model with Mask R-CNN [9] and the state-of-the-art method [6]. Finally, we show the representative results of each model. In order to ensure the accuracy of model performance evaluation, we apply five-fold cross-validation.

**Table 1.** Comparison of performance between our approaches (row C, row D, and row E) and other methods (row A, and row B)

|   | Models | DSC | Sensitivity | Specificity | HD |
|---|--------|-----|-------------|-------------|-----|
| A | Mukherjee et al [6] | 0.4288 | 0.4533 | 0.9961 | 59.6432 |
| B | Mask R-CNN [9] | 0.5989 | 0.6002 | **0.9998** | 62.0655 |
| C | B + H-RPN | 0.6508 | **0.7569** | 0.9996 | **49.0464** |
| D | B + H-RPN +LRS-Net | 0.6561 | 0.7126 | 0.9997 | 53.3620 |
| E | B + H-RPN + LRS-Net + DCRF | **0.6616** | 0.7069 | **0.9998** | 52.9130 |

Firstly, we reimplement the state-of-the-art method [6] (row A) and test the performance of Mask R-CNN [9] (row B) as the performance benchmark. As is shown in Table 1, the Mask R-CNN outperforms the state-of-the-art method in many indexes. Then we test the performance of our models by adding different modules step by step. It can be found that the equipment of H-RPN (row C) results in a significant increase in DSC, sensitivity, and HD. After adding the LRS-Net (row D), the DSC and specificity of the model are improved, which means the model is more conservative and achieve less FP. The area of background occupies the dominant position in MRI in the PWML segmentation task, so a small gain in specificity can represent a significant increase in background segmentation performance. Also, We get another small gain in DSC, specificity, and HD when we attach the DCRF to the end of the LRS-Net (row E). So, our final model is row E of Table 1.

**Table 2.** Expansion coefficient $k$ and model performance

| $k$ | 1.0 | 1.1 | 1.2 | 1.3 | 1.4 | 1.5 | 1.7 | 2 | 3 |
|-----|-----|-----|-----|-----|-----|-----|-----|---|---|
| DSC | 0.6326 | 0.6432 | 0.6472 | **0.6508** | 0.6497 | 0.6477 | 0.6460 | 0.6425 | 0.6366 |

When adding the surrounding information, we carry out a further experiment on the relationship between the expansion factor $k$ and the performance of our model based on row B of Table 1. $k = 1$ indicates that the ROI will not be expanded and the H-RPN will just be enhanced by the non-linear activation. As $k$ increases, the resources required for network training and inferencing will gradually increase. Therefore, we gradually train and test the performance of each model in the range of $k$ from 1 to 3 under the same condition. The results are shown in Table 2. When we enlarge the size of the ROI properly, the DSC of the model will increase to some extent, but the DSC will decrease if we continually expand the size of the ROI. So we enlarge the ROI by $k = 1.3$ to increase the performance of the model in the end.

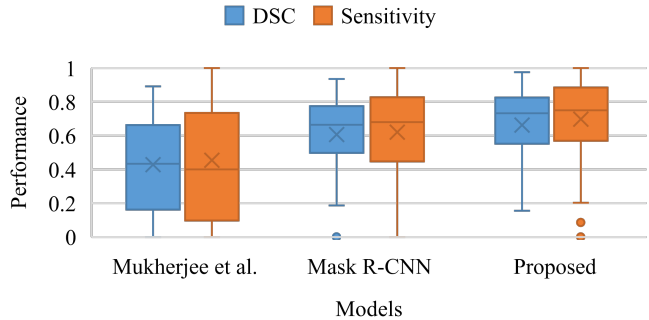

**Fig. 2.** Performance statistics of three methods on single slices

Secondly, we analyze the performance of our final model, Mask R-CNN [9], and the state-of-the-art method [6] on single MRI slices. The results are shown in Fig. 2. We find that the state-of-the-art method is not robust enough when there is a large amount of data, but the Mask R-CNN has a stable DSC and sensitivity in the segmentation of PWML. Compared with Mask R-CNN, our method has an improvement in the mean values of sensitivity and DSC. Also, the quartile spacing of sensitivity is smaller than that of Mask R-CNN. Which means that our method is more accurate, more stable, and more robust to segment different types of PWMLs.

Fig. 3 shows the segmentation results on the test set. We show the approximate area of the PWML in the red box in the original MRIs and make a difference display in the model segmentation results. Where the green pixel is true positive (TP) area, the red is the false positive (FP) area, and the blue is the false negative (FN) area. On the upper side is the number of the MRI section. For example, 8/77 is the 77th slice of the No.8 patient. In 8/77, there are some visible and low contrast lesions in the same slice. In 79/67, the lesion has a tiny area and a low contrast, but less surrounding interference information. In 82/67, the shape of the brain is unique. The following 83/48, 83/54 and 83/80 are images of different slices from a same patient. These images include lesions of different sizes, shapes, and contrasts. Compared with other algorithms, our method has a better performance and robustness in many cases.

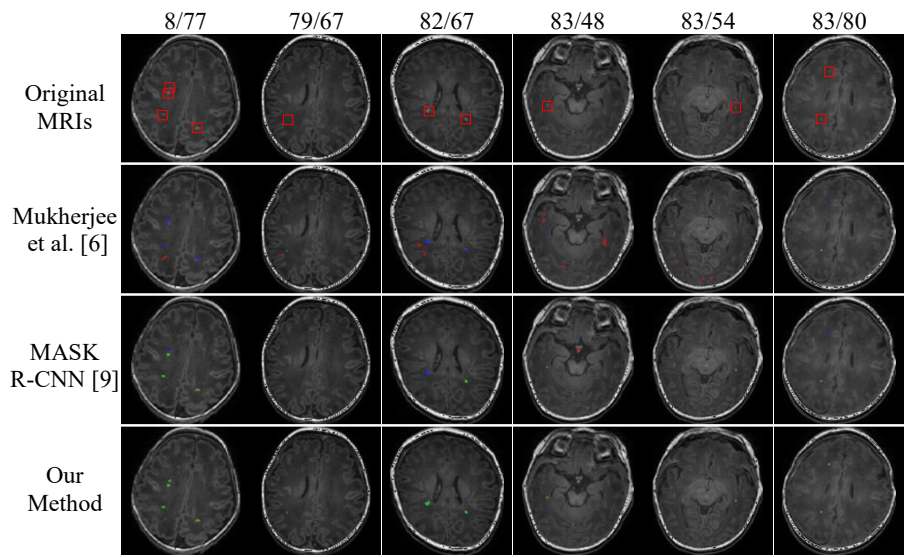

**Fig. 3.** Representative comparisons of different methods. PWMLs are located in red squares in the original MRIs. The Red pixel is FP, the blue is FN, and the green is TP in the results.

## 4    Discussion and Conclusion

In this paper, we propose an efficient deep learning model for PWML detection and segmentation. We use the non-linear activation function to enhance the H-RPN so that it can extract the region of interest more efficiently when there are many interference information and tiny targets. The heuristic method is used to enhance the H-RPN so that the surrounding information can guide the segmentation process and we can refine the class and the bounding box of the ROI. By this way, the small positioning deviation produced by H-RPN will not affect the performance of the segmentation process significantly, and the positive and negative samples of the partitioned network will be balanced. Therefore, our method has some benefits to other types of segmentation algorithms. Also, we design a lightweight segmentation network that can quickly and efficiently segment the PWML. We attach a DCRF to the end of the segmentation network to make the edge of the final segmentation more consistent with the texture in the original image. The result shows that the proposed method has a better performance and robustness compared with other existing methods in PWML segmentation.

In the future, we will further optimize the anti-interference performance of the model so that the model can separate PWML more effectively when there are some other pathological factors in MRIs.

## References

1. Tortora, D., Panara, V., Mattei, P. A., et al.: Comparing 3T T1-weighted sequences in identifying hyperintense punctate lesions in preterm neonates. American Journal of Neuroradiology 36(3), 581-586 (2015).
2. Kersbergen, K. J., Benders, M. J., Groenendaal, F., et al.: Different patterns of punctate white matter lesions in serially scanned preterm infants. PloS one 9(10), e108904 (2014).
3. Li, X., Gao, J., Wang, M., Zheng, J., Li, Y., Hui, E. S., Wan, M. & Yang, J.: Characterization of extensive microstructural variations associated with punctate white matter lesions in preterm neonates. American Journal of Neuroradiology 38(6), 1228-1234 (2017).
4. Cheng, I., et al.: White matter injury detection in neonatal MRI. Proceedings of the International Society for Optical Engineering. vol.8670, pp.86702L. SPIE, Florida (2013).
5. Cheng, I., Miller, S. P., Duerden, E. G., et al.: Stochastic process for white matter injury detection in preterm neonates. NeuroImage: Clinical 7, 622-630 (2015).
6. Mukherjee, S., Cheng, I., Miller, S., et al.: A fast segmentation-free fully automated approach to white matter injury detection in preterm infants. Medical & Biological Engineering & Computing 57(1), 71-87 (2019).
7. Milletari, F., Ahmadi, S. A., Kroll, C., et al.: Hough-CNN: deep learning for segmentation of deep brain regions in MRI and ultrasound. Computer Vision and Image Understanding 164, 92-102 (2017).
8. Ghafoorian, M., Mehrtash, A., Kapur, T., et al.: Transfer learning for domain adaptation in MRI: Application in brain lesion segmentation. In: International Conference on Medical Image Computing and Computer-Assisted Intervention, pp. 516-524. Springer, Cham (2017).
9. He, K., Gkioxari, G., Doll´ ar, P., Girshick, R.: Mask R-CNN. In: International Conference on Computer Vision, pp. 2980–2988 (2017).
10. Ren, S., He, K., Girshick, R., Sun, J.: Faster R-CNN: towards real-time object detection with region proposal networks. In: NIPS, pp. 91–99 (2015).
11. He, K., Zhang, X., Ren, S., Sun, J: Deep residual learning for image recognition. In: IEEE Computer Vision Pattern Recognition, pp. 770–778 (2016).
12. Lin, T. Y., Dollár, P., Girshick, R., He, K., Hariharan, B., Belongie, S.: Feature pyramid networks for object detection. In Proceedings of the IEEE Conference on Computer Vision and Pattern Recognition, pp. 2117-2125 (2017).
13. Kr¨ ahenb¨ uhl, P., Koltun, V.: Efficient inference in fully connected CRFs with gaussian edge potentials. In: NIPS, pp. 1–9 (2012).

14. Chen, L. C., Papandreou, G., Kokkinos, I., et al.: Deeplab: Semantic image segmentation with deep convolutional nets, atrous convolution, and fully connected crfs. IEEE Transactions on Pattern Analysis and Machine Intelligence 40(4), 834-848 (2018).
15. Kamnitsas, K., Ledig, C., Newcombe, V. F., et al.: Efficient multi-scale 3D CNN with fully connected CRF for accurate brain lesion segmentation. Medical Image Analysis 36, 61-78 (2017).
16. Snoek, J., Larochelle, H., Adams, R. P.: Practical bayesian optimization of machine learning algorithms. In Advances in neural information processing systems, pp. 2951-2959 (2012).